\documentclass[aps,twocolumn,nofootinbib]{revtex4-1}
\usepackage{amssymb}
\usepackage{amsmath}
\usepackage{graphicx}
\usepackage[sort&compress]{natbib}
\bibpunct[, ]{}{}{,}{s}{,}{,}

\usepackage[version=3]{mhchem} 
\begin{document}

\title{Equilibrium insertion of nanoscale objects into phospholipid
bilayers}

\author{Sergey Pogodin, Vladimir A. Baulin$^{\dag}$}
\affiliation{Departament d'Enginyeria Quimica, Universitat Rovira
i Virgili, Av. dels Paisos Catalans 26, 43007 Tarragona, Spain and
$^\dag $ICREA, Passeig Lluis Companys 23, 08010 Barcelona, Spain}

\email{vladimir.baulin@urv.cat}

\begin{abstract}
Certain membrane proteins, peptides, nanoparticles and nanotubes have rigid
structure and fixed shape. They are often viewed as spheres and cylinders
with certain surface properties. Single Chain Mean Field theory is used to
model the equilibrium insertion of nanoscale spheres and rods into the
phospholipid bilayer. The equilibrium structures and the resulting free
energies of the nano-objects in the bilayer allow to distinguish different
orientations in the bilayer and estimate the energy barrier of insertion.
\end{abstract}

\maketitle


\section{Introduction}

Biological membranes protect cells against the environment and
also provide for selective transport of nanoscale objects across
the membranes \cite{Alberts}. Peptides, nanotubes, nanoparticles
and other nano-objects with fixed shape can interact with
phospholipid bilayers \cite{Minko}, accumulate inside the
membranes \cite{Monticelli} and penetrate into
cells\cite{Biris,Porter1,Verma}. Surface activity of these objects
raises issues of biocompatibility and
cytotoxicity\cite{Hopfinger,Porter3}, which may limit their
biomedical applications. Understanding and controlling the
interactions of nanoscale objects with membranes is therefore a
key to the design of novel medical treatments and cell-active
substances which can modulate the thermodynamic properties of cell
membranes.

Although membrane active peptides, membrane proteins, nanotubes
and nanoparticles may significantly differ in composition and
chemical structure, the membrane activity implies the presence of
certain common features providing for such activity. Phospholipid
bilayers have alternating hydrophilic and hydrophobic regions
which are characterized by the corresponding thickness. Thus, the
size and the shape of nanoscale objects need to be compatible with
the bilayer structure to ensure their insertion
into the bilayer (\textit{e.g.} the concept of hydrophobic matching \cite%
{Smit}), cell penetration\cite{Kelley,Zorko} or pore formation\cite%
{Bayley,Carruthers}. The thickness of the hydrophobic core defines the
length scale for the size while the planar geometry of the bilayer defines
the template for the shape of nanoscale objects able to interact or insert
into phospholipid bilayers. The effect of the shape of nanoparticles interacting
with phospholipid bilayers have been studied in Ref. \cite{Ma}. The
nanoparticles-lipid bilayer interaction is determined by the contact area and the local
curvature of the nanoparticle at the contact point with the bilayer.
Thus, nanoscale objects of different nature can be viewed in the first
approximation as geometrical objects described only by the shape and the
surface properties\cite{Ma,Bloom,Hancock,Salonen,Klein4}.

\begin{figure}[tbp]
\includegraphics[width=8cm]{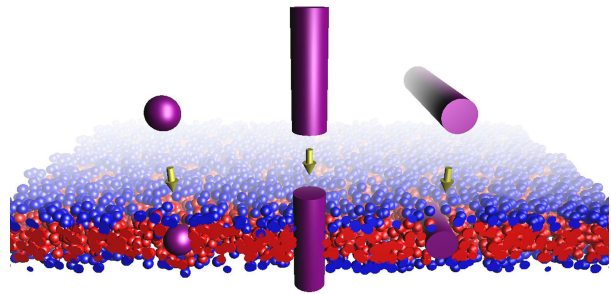}
\caption{Three geometries of nanoscale objects considered in the
text: insertion of a sphere (left), insertion of a
perpendicular cylinder (middle), and insertion of a parallel
cylinder (right) into the bilayer.} \label{Fig:scheme}
\end{figure}

Many theoretical studies of interaction of phospholipid bilayers with
nanoparticles and even peptides consider the simplest shapes: a sphere\cite%
{Monticelli,Bedrov,Qiao,Nielsen} and a cylinder\cite%
{Pogodin2,Sansom1,Huang,May,Klein4,DesernoPore} with uniform or
patterned surface properties. This allows to describe a broad
range of objects of similar size within a single concept and
provide a general picture on the mechanism and the molecular
details of interaction with the bilayers. Coarse-grained Molecular
Dynamics (MD) provides the most detailed information about the
molecular structure and the dynamics of interaction of nanoscale
objects with phospholipid
bilayers\cite{Monticelli,Ma,Gu,Bedrov,Sansom1,Klein4,Klein3,Klein6,Sarkisov1,Sarkisov2}.
However, this method becomes computationally expensive and time
consuming when the size of the system and the number of
interacting molecules increases\cite{Muller2}. Calculation of
equilibrium properties and equilibrium free energies requires
equilibration process which may exceed microseconds\cite{Muller2}.
One of the alternatives is the use of the self-consistent field
theories, where the equilibrium structure of the bilayer is found
from the solution of the self-consistency equations. Recently,
Single Chain Mean Field (SCMF) theory has been applied for
modeling of the phospholipid bilayers at different levels of
coarse-graining\cite{Pogodin}. It was shown, that essential
equilibrium properties of the phospholipid bilayer such as the
thickness, position of different groups, the compressibility
constant and the area per lipid, can be reproduced with high
accuracy. High speed of calculations is obtained by the decoupling
of the correlations between the molecules and the fluctuations,
taking into account the symmetries in the system. In this manner,
the multi-molecule problem is reduced to a single molecule in the
external self-consistent field problem. The molecule conformations
and the fields are found from the self-consistency condition and
gives the free energy of the equilibrium structures as a result of
calculations. This method has been applied to estimate the free
energies of insertion of a carbon nanotube into a phospholipid
bilayer \cite{Pogodin2} and to elucidate the patterning of a
carbon nanotube for physical translocation through the
bilayer\cite{Pogodin3}.

In this publication, we first present equations of the SCMF theory
for the interaction of phospholipid bilayer and nanoscale objects.
This theory is then applied for calculation of equilibrium
structures and free energies of resulting structures of
phospholipids and nanoscale objects, modeled as spheres and
cylinders with given surface properties (Fig. \ref{Fig:scheme}).

\section{Theory}

\begin{figure}[tbp]
\includegraphics[width=8cm]{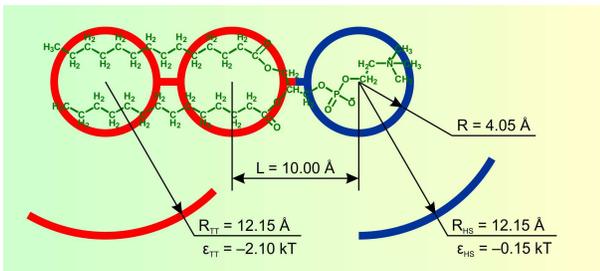}
\caption{The 3-beads model of the DMPC phospholipid molecule. $R$
is the radius of the hydrophobic T-beads (red) and hydrophilic
H-bead (blue), $L$ is the bonds length, $R_{TT}$ and $R_{HS}$ are
the interaction radii of the corresponding stepwise attractive
potentials.} \label{Fig:lipid}
\end{figure}

The SCMF theory uses a coarse-grained representation of a
phospholipid molecule as a set of spherical beads interacting with
the fields through simple effective potentials \cite{Pogodin}. The
simplest 3-beads model (Fig. \ref{Fig:lipid}) is able to reproduce
with high accuracy the equilibrium structure and properties of
DMPC phospholipid bilayer, such as the average interfacial area
per lipid, the thickness of the bilayer and the core, and the
stretching modulus \cite{Pogodin}. In this model the phospholipid
molecule is represented by two hydrophobic beads (T) and one
hydrophilic bead (H) of the same radius 4.05 \AA, joined
consequently by rigid bonds of length 10.0 \AA, and able to bend
around the central T-bead. Two T-beads, belonging to different
molecules, interact with energy gain $\epsilon_{TT}= -2.10$ kT at
distances between their centers smaller than 12.15 \AA.
Additionally, the H-beads interact with implicit solvent
molecules, assumed as well to be represented by spherical beads of
the same radius 4.05 \AA, with energy gain $\epsilon_{HS}=-0.15$
kT at the distances closer than 12.15 \AA.

The free energy of a system, containing $N$ lipid molecules and
the solvent, depends in general on the total probability
distribution function $\rho $ of the system. In the mean field
approximation one can split this multi-molecule probability
distribution function into a product of the corresponding
single-molecule distribution functions. This allows to write the
total free energy of the 3-beads model as a sum
\begin{equation}
F=-S_{L}-S_{S}+U_{0}+U_{TT}+U_{HS}+U_{steric}  \label{F}
\end{equation}%
where $S_{L}$ and $S_{S}$ are the entropy contributions of the lipids and
the solvent, $U_{0}$ describes the intra-molecular and position-dependent
energies of the lipids, $U_{TT}$ and $U_{HS}$ represent the inter-molecular
interactions between the pairs of T-beads and between H-beads and the
solvent, and the last term, $U_{steric}$, takes into account the hard-core
steric repulsions between the molecules.

The entropy part of the free energy is written as
\begin{equation}
-S_{L}-S_{S}=N\left\langle \ln \left[ \rho (\gamma )N\Lambda \right]
\right\rangle +\int c_{S}(\pmb r)\ln \left[ c_{S}(\pmb r)\Lambda _{S}\right]
d\pmb r  \label{SL&SS}
\end{equation}%
where $\rho (\gamma )$ is the density of probability of a single lipid
molecule to be in the conformation $\gamma $ (along with relative
positions of the beads in the molecule with respect to each other,
the conformations also include their spatial position and
orientation in the system), $c_{S}(\pmb r)$ is the solvent
concentration at the point $\pmb r$, $\Lambda $ and $\Lambda _{S}
$ are the de Broglie lengths for the lipids and the solvent respectively, which shift the
absolute value of the free-energy but have no effect on the final
results. The angular brackets denote the average over $\rho
(\gamma )$, and the integral is taken over all the accessible
volume of the system.

The intra-molecular interaction energy $U_{0}$ is simply the average over $%
\rho (\gamma )$
\begin{equation}
U_{0}=N\left\langle U_{0}(\gamma )\right\rangle
\end{equation}%
where $U_{0}(\gamma )$ is the internal energy of the lipid molecule in the
conformation $\gamma $. It can include interactions between the beads
composing the molecules and interaction with the external fields and the
walls (or impermeable immobile objects).

The combined inter-molecular energy terms are written as
\begin{eqnarray}
U_{TT}+U_{HS}&=&\frac{N(N-1)}{2}\int \left\langle u_{T}(\gamma ,\pmb %
r)\right\rangle \left\langle c_{T}(\gamma ,\pmb r)\right\rangle d\pmb %
r \\ \nonumber
 &&+N\int \left\langle u_{S}(\gamma ,\pmb
r)\right\rangle c_{S}(\pmb r)d\pmb r \label{HTTHHS}
\end{eqnarray}%
where $u_{T}(\gamma ,\pmb r)$ and $u_{S}(\gamma ,\pmb r)$ are the
interaction potentials for T-beads and the solvent at the point $\pmb r$,
created by the lipid molecule in the conformation $\gamma $, and $%
c_{T}(\gamma ,\pmb r)$ is the concentration of T-beads of the lipid in
conformation $\gamma $ at the point $\pmb r$.

\ref{HTTHHS} is valid only if the potentials $u_{T}(\gamma
,\pmb r)$ and $u_{S}(\gamma ,\pmb r)$ are finite and smooth functions of $%
\pmb r$ and therefore cannot contain any hard-core steric repulsion
interactions, which are included in a separate term, $U_{steric}$,
representing the incompressibility condition at every point $\pmb r$ of the
system
\begin{equation}
U_{steric}=\int \lambda (\pmb r)\left( \phi _{0}-v_{S}c_{S}(\pmb %
r)-N\left\langle \phi (\gamma ,\pmb r)\right\rangle \right) d\pmb r
\label{Hsteric}
\end{equation}
Here $\lambda (\pmb r)$ is the Lagrange multiplier and $\phi _{0}$ is the
total volume fraction occupied by the lipids and the solvent ($0<\phi _{0}<1$%
), $v_{S}$ denotes the volume of the solvent molecule, and $\phi (\gamma ,%
\pmb r)$ is the volume fraction at the point $\pmb r$ occupied by the lipid
in the conformation $\gamma $.

To find the equilibrium distribution, one should minimize the
total free energy (\ref{F}) with respect to $\rho (\gamma )$ and
the solvent distribution $c_{S}(\pmb r)$ subject to the
incompressibility condition (\ref{Hsteric}). The minimization
leads to the following set of integral equations

\begin{eqnarray}
\rho (\xi ) &=&\frac{1}{Z}\exp \left[ -U_{0}(\xi )-(N-1)\int u_{T}(\xi ,\pmb %
r)\left\langle c_{T}(\gamma ,\pmb r)\right\rangle d\pmb r-\right.  \notag \\
&&\left. \int u_{S}(\xi ,\pmb r)c_{S}(\pmb r)d\pmb r+\int \lambda (\pmb %
r)\phi (\xi ,\pmb r)d\pmb r\right]  \label{vrho}
\end{eqnarray}

\begin{equation}
v_{S}\lambda (\pmb r)=\ln \left[ v_{S}c_{S}(\pmb r)\right]
+N\left\langle u_{S}(\gamma ,\pmb r)\right\rangle \label{vS}
\end{equation}

\begin{equation}
v_{S}c_{S}(\pmb r)=\phi _{S}(\pmb r)=\phi _{0}-N\left\langle \phi (\gamma ,%
\pmb r)\right\rangle  \label{cS}
\end{equation}
where the normalization factor $Z$ ensures that $\int \rho (\gamma
)d\gamma=1$. These equations can be solved numerically, if the integrals
and the averages are replaced by appropriate sums, which reduce
the problem to solution of algebraic equations. To discretize
the space, the simulation box is divided into a set of auxiliary
cells, where all spatial dependent variables have a constant
value. The division into cells should also take into account the
symmetry of the solution, which provides the speed up of the
calculations. For example, if the lipid system is expected to
self-organize into a flat bilayer, the division of the simulation
box according to 1D geometry into parallel slits results in
considerably smaller number of cells than the 3D division into a
cubic grid with the same resolution. The conformational space
$\gamma $ of a single lipid is sampled with the Rosenbluth
algorithm\cite{Rosenbluth}, the conformations are placed in the
box and the spatial distributions $u_{T}(\gamma ,\pmb r)$, $u_{S}(\gamma ,%
\pmb r)$, $c_{T}(\gamma ,\pmb r)$, $\phi (\gamma ,\pmb r)$ as well
as other necessary conformational-dependent properties are
calculated. Finally, the averages of the fields $f$ in the
equations \ref{vrho}-\ref{cS} are replaced by the sums

\begin{equation}
\left\langle f(\gamma )\right\rangle =\frac{1}{\kappa }\sum_{\gamma }\frac{%
\rho _{\gamma }f_{\gamma }}{w_{\gamma }}
\end{equation}%
where $f_{\gamma }$ is the value of this field, and $\rho _{\gamma
}$ is the probability of the conformation $\gamma $ belonging to
the generated sampling (thus $\rho_\gamma$ is determined only over
the sampling, while $\rho(\gamma)$ used above was determined in
the full conformational space of the molecule). $w_{\gamma }$ is
the value of the
Rosenbluth weight (\textit{i.e.} biased probability of the conformation $%
\gamma $ is introduced into the sampling during the generation).
$\kappa $ is the size of the conformational sampling. Similarly,
the integrals over the space are replaced by the sums as

\begin{equation}
\int f(\pmb r)d\pmb r=\sum_{i}f_{i}V_{i}  \label{discretization}
\end{equation}
where $f_{i}$ is the value of the field in the $i$-th auxiliary
cell of the simulation box and $V_{i}$ is the volume of the $i$-th
cell. As a result, the system of integral equations
\ref{vrho}-\ref{cS} is reduced to the closed set of algebraic
equations of finite but huge number $\kappa $ of unknown variables
$\rho _{\gamma }$. Since the larger $\kappa $, the more accurate
the solution is, the practical choice of $\kappa $ which provide a
reasonable accuracy is limited by the \textit{representative}
sampling. The easiest way to check the accuracy is to perform
several simulations with independently generated random samplings
and compare their results. Thus we start our simulations with some
small value of $\kappa$, and then perform series of similar
simulations for larger $\kappa$ till the moment when further
increase of $\kappa$ will not lead to significant change of
calculated results. For the simulations reported in this work we
used $\kappa = 750 000$, which provide accuracy of the calculated
energies about $\pm 10$ kT.

The number of auxiliary cells in the box should be high enough to
provide acceptable spatial resolution, but if it is relatively
small one can decrease the number of independent variables by
introducing new independent variables, average concentration of
T-beads, $c_{T,i}\equiv \left\langle
c_{T}(\gamma ,\pmb r_{i})\right\rangle $, potential of the solvent, $%
u_{S,i}\equiv \left\langle u_{S}(\gamma ,\pmb r_{i})\right\rangle
$, and the volume fraction occupied by lipids, $\phi _{i}\equiv
\left\langle \phi (\gamma ,\pmb r_{i})\right\rangle $. Depending
on the method of solution of equations\footnote{According to Ref.
\cite{Pogodin}, the independent variables are $c_{T,i}$, $c_{H,i}$
and $\phi _{i}$. One can show that the expression $N\left\langle
u_{S}(\gamma ,\pmb r)\right\rangle $, which is the average
potential experienced by the solvent molecule at the point $\pmb
r$, is approximately equal to $N\epsilon _{S}\left\langle
c_{H}(\gamma ,\pmb r)\right\rangle $, where $\epsilon _{S}$ is the
integral of the H-S interaction potential around a solvent
molecule. One can easily replace the variables $u_{S,i}$ by
$c_{H,i}$. This was indeed necessary in the general case,
considered in the previous article, but in this context it can be
avoided for the sake of simplicity.}, such reduction of the number
of independent variables can speed up calculations.

The equations of the SCMF theory are written for the canonical
(NVT) ensemble. However, the simulation of the phospholipid
bilayer with inserted nanoscale object would require the
grand-canonical ensemble, since the inserted object may expel the
phospholipids out of the simulation box. Fortunately, the SCMF
technique is fast enough to perform series of canonical
calculations for a part of bilayer with variable number of lipids
$N$ in the simulation box. One can consider then the simulation
box as an open part of a larger closed system and thus estimate
the free energy $F^{\ast }$ of the whole system as a function of
$N$. The equilibrium value of $N$ will correspond to the minimum
of $F^{\ast }$, thus series of canonical SCMF calculations are
equivalent to a grand-canonical calculation. In the simulations
performed in the present work (size of the system about
$100.0\times 100.0\times 60.0$ \AA, two-dimensional geometry with
spatial resolution 2.0 \AA, and sampling size $\kappa = 750000$),
calculation of one energy point in the grand-canonical ensemble
with current version of our SCMF code, demanded, on average, about
one day on a 32-cores machine. Full three-dimensional simulations
should be significantly slower, but we believe that further
improvements of our code will significantly speed-up its'
performance, thus making possible three-dimensional simulations in
a reasonable time.

\section{Results and discussion}

The equations of the previous section allow to calculate the
equilibrium properties and the free energy of the phospholipid
bilayer and its interaction with nanoscale objects with a fixed
shape. Simulation of a phospholipid bilayer is a one dimensional
problem if we neglect the bilayer bending. Since the free energy
of a homogenous bilayer is proportional to the area (or the number
of lipids in the bilayer) and the interfacial area per lipid and
the bilayer thickness are constant along the bilayer, the free
energy of a large system $F^{\ast }$ can be written as
\begin{equation}
F^{\ast }=V^{\ast }f_{S}+N^{\ast }f(A),\quad f(A)=\frac{F(V,N,A)-Vf_{S}}{N}%
,\quad f_{S}=\frac{\phi _{0}}{v_{S}}\ln \frac{\phi _{0}}{v_{S}}
\label{FSysPureFlatMembrane}
\end{equation}%
where $V^{\ast }$ is the volume and $N^{\ast }$ is the total
number of lipids of the large system, $f_{S}$ is the free energy
density of the region occupied by the pure solvent, and $f(A)$ is
the free energy per lipid of the bilayer, which depends on the
average surface area per lipid $A$.

\begin{figure}[tbp]
\includegraphics[width=8cm]{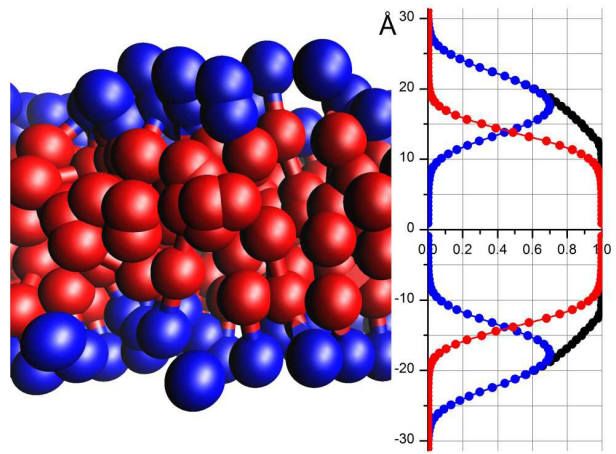}
\caption{Mean field snapshot of DMPC bilayer at equilibrium within
the 3-beads model (left), and the corresponding concentration
profiles (right) of hydrophobic (red), hydrophilic (blue) and
total (black) volume fractions of the beads.} \label{Fig:bilayer}
\end{figure}

The result of the SCMF simulation is the free energy $F(V,N,A)$ of
the box of the total volume $V$ with the number of lipids $N$.
These expressions allow to construct the free energy per lipid
$f(A)$. The equilibrium area per lipid $A_{0}$, which minimizes
the total free energy of a large system $F^{\ast }$, can be found
by the minimization of $f(A)$, obtained from the series of SCMF
simulations, while the simulation corresponding to $A_{0}$ gives
the equilibrium distribution of T- and H-beads inside the bilayer
and the equilibrium thickness (Fig. \ref{Fig:bilayer}).

Similar approach is applied to study the insertion of a rigid
spherical or cylindrical nano-object into the phospholipid
bilayer. A part of the object located inside the simulation box is
represented by a region prohibited for lipids and the solvent to
enter, and the surface of the object can interact preferentially
with phospholipid beads. This restricted area is taken into
account during the sampling generation by the additional bias and
the corresponding contribution to the intra-molecular energy
$U_0(\gamma)$. In the case of a spherical object and a cylindrical
object oriented perpendicular to the bilayer one can use the
two-dimensional cylindrical geometry with the symmetry axis
passing through the center of the sphere (cylinder) perpendicular
to the bilayer plane. The simulation of a cylindrical object in
the parallel orientation has been done within two-dimensional grid geometry
\textit{i.e.} assuming infinite cylinder. To limit the bending of
the bilayer, which permits the inserted object to escape from the
bilayer, the hard walls on the top and bottom of the simulation
box have been introduced.

Assuming that a part of the membrane outside the simulation box is
not perturbed by the insertion of the object and, thus, is in
equilibrium state, we estimate the total free energy change due to
the insertion of the object to a given position $\pmb p$ in the
box as

\begin{equation}  \label{DeltaF}
\Delta F^* = F(V,N,\pmb p) - Nf(A_0) - (V-V(\pmb p))f_S
\end{equation}
where $F(V,N,\pmb p)$ is the free energy of the box and $V(\pmb
p)$ is the volume occupied by the object inside the box.

\begin{figure*}[tbp]
\includegraphics[width=15cm]{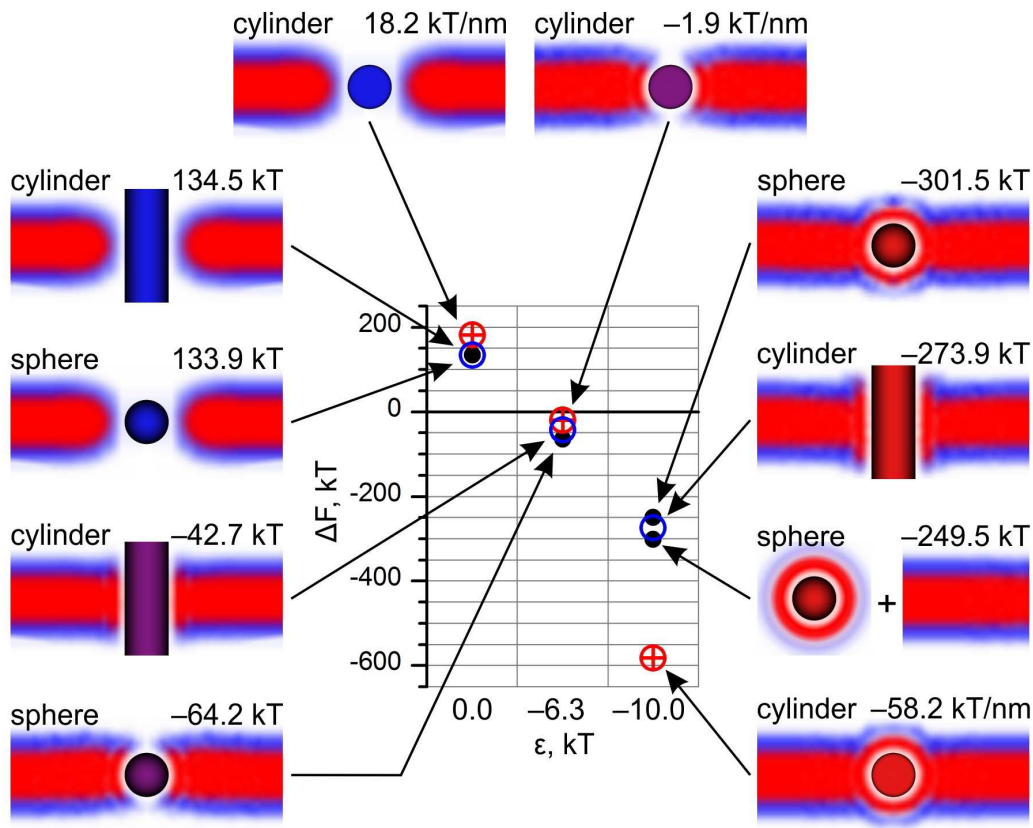}
\caption{Snapshots of the equilibrium insertion of a spherical and
a cylindrical object into the DMPC bilayer. The color of the
objects in the insets illustrates the hydrophobicity level
(interaction with T-beads): $\epsilon = 0.0$ (blue), $\epsilon =
-6.3$ kT (purple), $\epsilon=-10.0$ kT (red). Parallel orientation
of a cylinder (denoted by the energy/nm), corresponds to the
cylinder of length $10.0$ nm.} \label{Fig:diagram}
\end{figure*}

The results of the free energy calculations for spheres and
cylinders of diameter $2.43$ nm with homogeneous surface
properties are summarized in Fig. \ref{Fig:diagram}. The surface
properties are described by a single interaction parameter between
the T-beads and the surface, $\epsilon$. One can clearly see, that
the free energy changes due to the insertion of a sphere, or a
cylinder in perpendicular orientation which has a similar contact
area with lipids, have relatively close values for all considered
levels of hydrophobicity $\epsilon$. Thus, the shape of the object
is not so important as the contact area with lipids. This
conclusion is consistent with the previous observations\cite{Ma}.
In contrast, the cylinder in the parallel orientation has a larger
contact area with phospholipid tails and disturbs much larger part
of the bilayer. Thus, the parallel orientation has sufficiently
lower equilibrium free energy than the perpendicular orientation
(the points in the plot correspond to a cylinder with the length
$10.0$ nm, while the numbers in the insets correspond to the
energy per nm).

Hydrophilic and hydrophobic objects induce different rearrangement
of lipids in the bilayer (see the insets of Fig.
\ref{Fig:diagram}). Hydrophilic objects of similar size create
pores of similar size with the heads of the lipids oriented inside
the pores. Hydrophobic objects attract the tails of phospholipids
and they tend to come closer to the surface. For most hydrophobic
objects, the meniscus provoked by the wetting of the hydrophobic
object can be formed. For intermediate interaction parameters, the
pore is not complete and is combined with the partial wetting at
the edges, thus provoking the thinning of the bilayer in contact
with the nano-object. This diagram also suggest the preferential
orientation of cylindrical objects in the bilayer. Hydrophilic
cylinder has lower free energy in perpendicular orientation, while
the most hydrophobic cylinder has lower free energy in parallel
orientation. For an intermediate hydrophobicity, there is a
transition in orientation from perpendicular to parallel
orientation which depends also on the length of the cylinder. For
$\epsilon = -6.3$ kT the transition from perpendicular to parallel
orientation occurs at the length of the cylinder 10 nm, where both
orientations have almost the same energies. Longer cylinders would
favor parallel orientation.

The uptake of a hydrophobic particle into the core of a bilayer is
not the only equilibrium solution of the equations. There exists a
solution which corresponds to a hydrophobic particle covered by
phospholipids floating around and coexisting with a bilayer (third
snapshot from the top in the right column of Fig.
\ref{Fig:diagram}). We have compared the free energy of such
structure with the energy of the inserted sphere into the bilayer
in the most hydrophobic case, $\epsilon=-10.0$ kT. The bilayer
with the incorporated sphere has a lower energy, while the free
energy difference with the sphere in the solution is of the order
$50.0$ kT, indicating that the hydrophobic sphere would unlikely
escape. Similar conclusion can be drawn for the case of
cylindrical particles lying parallel to the bilayer.

\section{Conclusions}

We have developed the SCMF theory for the interaction of nanoscale
objects with phospholipid bilayers. This numerical method provides
for detailed microscopic information about the structure of
phospholipid bilayers and their essential equilibrium thermodynamic
properties. The method gives the total equilibrium free energy of
self-assembled structures as the output of calculations, which
allows for direct comparison of different equilibrium structures
involving nanoscale objects and phospholipid bilayers.

We have applied the SCMF theory to investigate the interaction of
phospholipid bilayer with cylindrical and spherical nanoparticles
of diameter 2.43 nm and surface properties ranging from
hydrophilic to strongly hydrophobic. We have shown that the shape
of the nanoparticles is not as important as the contact area with
the lipids. Hydrophobic and hydrophilic objects induce different
reorganization of lipids around the nanoparticles. More complex
nano-objects will be considered in future works.

\begin{acknowledgements}

The authors acknowledge the UK Royal Society International Joint
Project with Cambridge University and Spanish Ministry of
education MICINN project CTQ2008-06469/PPQ.

\end{acknowledgements}


\providecommand*\mcitethebibliography{\thebibliography} \csname
@ifundefined\endcsname{endmcitethebibliography}
  {\let\endmcitethebibliography\endthebibliography}{}

\end{document}